\documentclass[12pt]{article}

\usepackage[reqno]{amsmath}
\usepackage{mathrsfs}
\usepackage{amssymb}

\usepackage{rotating} %for landscape format

\usepackage{bbm}
\usepackage{epsfig}
\usepackage{array}
\usepackage{float}
\usepackage{color}
\usepackage{graphicx}% Include figure files
\usepackage{feynmf}

\usepackage{a4}
\usepackage{a4wide}
\usepackage{wasysym}
\def\gs{\mathrel{
   \rlap{\raise 0.511ex \hbox{$>$}}{\lower 0.511ex \hbox{$\sim$}}}}
\def\ls{\mathrel{
   \rlap{\raise 0.511ex \hbox{$<$}}{\lower 0.511ex \hbox{$\sim$}}}}

\newcommand{\be}{\begin{eqnarray}}
\newcommand{\ee}{\end{eqnarray}}

\begin{document}

%For feynmf-Package:
\setlength{\unitlength}{1mm}

\begin{titlepage}
\title{\vspace*{-2.0cm}
%\hfill {\small hep--ph/xxxxxx}\\[20mm]
\bf\Large
General Conditions for Lepton Flavour Violation at Tree- and 1-Loop Level
\\[5mm]\ }

\author{
Alexander Blum\thanks{email: \tt alexander$.$blum@mpi-hd.mpg.de}~~,~~
Alexander Merle\thanks{email: \tt alexander$.$merle@mpi-hd.mpg.de}
\\ \\
{\normalsize \it Max-Planck-Institut f\"ur Kernphysik,}\\
{\normalsize \it Postfach 10 39 80, 69029 Heidelberg, Germany}
}
\date{\today}
\maketitle
\thispagestyle{empty}

\begin{abstract}
\noindent
In this work, we compile the necessary and sufficient conditions a theory has to fulfill in order to ensure general lepton flavour conservation, in the spirit of the Glashow-Weinberg criteria for the absence of flavour-changing neutral currents. At tree-level, interactions involving electrically neutral and doubly charged bosons are investigated. We also investigate flavour changes at 1-loop level. In all cases we find that the essential theoretical requirements can be reduced to a few basic conditions on the particle content and the coupling matrices. For 1-loop diagrams, we also investigate how exactly a GIM-suppression can occur that will strongly reduce the rates of lepton flavour violating effects even if they are in principle present in a certain theory. In all chapters, we apply our criteria to several models which can in general induce lepton flavour violation, e.g.\ $LR$-symmetric models or the MSSM. In the end we give a summarizing table of the obtained results, thereby demonstrating the applicability of our criteria to a large class of models beyond the Standard Model.
\end{abstract}

\end{titlepage}

%%%%%%%%%%%%%%%%%%%%%%%%%%%%%%%%%%%%%%%%%%%%%%%%%%%%%%%%%%%%%%%%%%%%%%
\section{\label{sec:intro} Introduction}
%%%%%%%%%%%%%%%%%%%%%%%%%%%%%%%%%%%%%%%%%%%%%%%%%%%%%%%%%%%%%%%%%%%%%%

In the last decades, it has been shown that the Standard Model (SM) of elementary particle physics is an excellent description of physics up to the energy scales, we have been able to probe so far. Furthermore, it has predicted several particles which then indeed have been discovered, among them the neutral $Z$-boson~\cite{Arnison:1983mk} and the $t$-quark~\cite{Abachi:1995iq}. The only missing building block is the predicted neutral Higgs boson, but even that is expected to be discovered in the near future at high-energy particle colliders such as the upcoming LHC experiment~\cite{LHC}.\\
However, the SM also has several problems and cannot explain all phenomena: E.g.\ it provides no candidate particle for the observed Dark Matter in the Universe~\cite{Bertone:2004pz}, it has no explanation for the obvious baryon asymmetry~\cite{Riotto:1998bt}, and provides no mechanism for stabilizing the Higgs mass against radiative corrections~\cite{Martin:1997ns}. Apart from the gauge symmetries, the SM also conserves several quantum numbers more or less by accident, among them lepton number and furthermore lepton flavour. As the conservation of lepton flavour is not in fact an integral part of the SM, it is often lost in popular extensions, designed to address, among others, the problems mentioned above. They then predict lepton flavour violating processes such as \ $\mu \rightarrow e \gamma$ or $\mu^- \rightarrow e^- e^- e^+$ (which are perfectly allowed by energy and charge conservation). Because of this, impressive experimental activities have been undertaken to detect such processes: Some current experimental limits are $BR(\mu \rightarrow e \gamma)<1.2 \cdot 10^{-11}$ (MEGA experiment, Ref.~\cite{Brooks:1999pu}), $BR(\mu \rightarrow 3e)<1.0 \cdot 10^{-12}$ (SINDRUM experiment, Ref.~\cite{Bellgardt:1987du}), or $BR(\mu {\rm Ti}\rightarrow e {\rm Ti})<4.3 \cdot 10^{-12}$ (SINDRUM II experiment, Ref.~\cite{Dohmen:1993mp}).\footnote{For a nice collection of further experimental bounds on various lepton flavour violating processes consult Ref.~\cite{Blanke:2007db}.} The bound for the first branching ratio will be improved in the near future: The upcoming MEG experiment is expected to reach a sensitivity of $1.2\cdot 10^{-13}$ at $90\%$~C.L.\ and a single event sensitivity of even $3.7\cdot 10^{-14}$~\cite{Ritt:2006cg}.\\
So far, however, only upper limits for the branching ratios of these processes can be given. If one tries to parameterize the bounds for their rates, the corresponding numerical coefficients are already quite small \cite{deGouvea:2007xp}. Especially if MEG does not observe any lepton flavour violating decays, this leads to the question, whether lepton flavour conservation (at least at the tree- or 1-loop level) needs to be imposed as a general condition on extensions of the SM. It is our aim in this paper to give such criteria, i.e.\ to determine sufficient and necessary conditions for the conservation of lepton flavour in a general theory, which incorporates the SM. By giving necessary conditions for lepton flavour conservation, our results can also be applied if MEG does in fact observe lepton flavour violating decays: As lepton flavour violation occurs in may extensions of the SM, no single theory can be considered to be proven by such results. By applying the criteria developed in this work, one can determine what exactly is necessary to obtain lepton flavour violating processes, and what a minimal lepton flavour violating extension of the SM must contain.\\
Actually, lepton flavour violating processes have already been observed, in neutrino oscillation experiments (see e.g.\ Ref.~\cite{Barger:2003qi}). However, even if one incorporates these results into the SM, by allowing for massive neutrinos and off-diagonal elements in the leptonic mixing matrix $U_{\rm PMNS}$, the most optimistic prediction (using a neutrino mass of $\sim 1$~eV) for $BR(\mu \rightarrow e \gamma)$ will roughly be $10^{-47}$ \cite{Cheng:1985bj}, making a detection impossible. This is because there is another mechanism at work here, the GIM-mechanism, well-known from flavour changing processes among quarks. So, when considering whether lepton flavour (among charged leptons) is conserved in a given theory, we also consider the possibility of lepton flavour being violated but all relevant processes being GIM-suppressed, as this is the most efficient known mechanism for suppressing such decays.\\
The groundbreaking work on Flavour Changing Neutral Currents (FCNCs) was done by Glashow and Weinberg~\cite{Glashow:1976nt}, already in the late 70's. This paper only dealt with flavour violation in the quark sector, and of course at that time, it was e.g.\ not known how many quark flavours indeed exist in our world, and the exact structure of the weak interaction was unknown as well. In the meantime we believe to know these things better, so it is worth reconsidering such criteria for flavour violation, which we do herewith for the leptonic sector, in the framework of more recent knowledge. Our basic idea is that we give general conditions necessary for lepton flavour violation (LFV) not to occur. If these conditions are not fulfilled it will -- in general -- be possible to have LFV-processes, assuming that there are no accidental cancellations or further suppressions in the theory. Many of our results are known, or at least often used implicitly, however, no concise overview of these criteria, akin to the work of Glashow and Weinberg, is currently available.\\
We will only investigate renormalizable interactions and do not consider higher-dimensional operators, since in a non-renormalizable theory explicit lepton flavour violating operators, such as
\begin{equation}
 \frac{1}{\Lambda^2} (\overline{\mu} e)(\overline{e} e)
 \label{eq:explicit}
\end{equation}
where $\Lambda$ is the energy scale at which lepton flavour is violated, can be simply added to the Lagrangian. In addition to the general criteria, we investigate in each section several examples and use our general criteria to give concrete conditions for the parameters in the respective models.\\
We start in Sec.~\ref{sec:FCNCs} with lepton Flavour Changing Neutral Currents (FCNCs) and present the criteria for the absence of lepton FCNCs for scalar and vector bosons as mediators of a flavour change (FC). These criteria can also be applied to quarks. In Sec.~\ref{sec:2CC}, we then turn to doubly charged exchange bosons. Such particles do not appear in the SM but naturally arise in several theories beyond the Standard Model (BSM theories). After that, in Sec.~\ref{sec:1-loop}, we show that it is also easy to find general criteria for the absence or occurrence of lepton flavour violation at 1-loop level and for a possible GIM-suppression. In each section we give some examples by investigating concrete models in which our general conditions turn out to be applicable. We finally give a summary of our results and conclude in Sec.~\ref{sec:Summary}. Notations and conventions we have used are listed in the Appendix.

%%%%%%%%%%%%%%%%%%%%%%%%%%%%%%%%%%%%%%%%%%%%%%%%%%%%%%%%%%%%%%%%%%%%%%
\section{\label{sec:FCNCs} Lepton flavour changing neutral currents at tree-level}
%%%%%%%%%%%%%%%%%%%%%%%%%%%%%%%%%%%%%%%%%%%%%%%%%%%%%%%%%%%%%%%%%%%%%%

In general, a neutral current interaction that changes the flavour of a fermion $f_i$ (we here speak of general fermions, as the results of this general section can also be applied to quarks) can be mediated by a neutral scalar or a neutral vector boson that couples to a fermion $f_i$ as well as to a fermion $f_j$ with a different flavour index $j\neq i$. Writing down the most general Lagrangians for both cases, the scalar interaction looks like
\begin{equation}
 \mathcal{L}_{\rm scalar}= S \overline{f} \left( C_L \mathcal{P}_L + C_R \mathcal{P}_R \right)f + h.c.,
 \label{eq:scalarint}
\end{equation}
and the vector interaction has the form
\begin{equation}
 \mathcal{L}_{\rm vector}= V_{\mu} \overline{f} \gamma^{\mu} \left( C_L \mathcal{P}_L + C_R \mathcal{P}_R \right)f + h.c.,
 \label{eq:vectorint}
\end{equation}
where $f=(f_1,f_2,...,f_N)^T$ is a vector and $C_L$ and $C_R$ are numerical coefficients (matrices), all in flavour space.

Following the procedure given in the Appendix, the vector $f$ can be rotated from the interaction eigenstate into the mass eigenstate $f'$ via a transformation matrix $U$ according to $f=U f'$ which is, in general, not the same for left- and right-handed fermions. Now the question is, how the interaction terms Eqs.~\eqref{eq:scalarint} \&~\eqref{eq:vectorint} look like if one transforms the interaction eigenstates $f$ into the corresponding mass eigenstates $f'$:

\begin{enumerate}

\item[S)] $S$ is a neutral scalar by assumption, hence we can define it as real by absorbing any phase in the coupling matrices. In the flavour space vector notation, the scalar interaction as written in Eq.~\eqref{eq:scalarint} can be simplified giving
\begin{eqnarray}
 \mathcal{L}_{\rm scalar} &=& S \overline{f} \left( C_L \mathcal{P}_L + C_R \mathcal{P}_R \right)f + h.c.=\nonumber \\
 &=& S (\overline{f_R} C_L f_L + \overline{f_L} C_R f_R ) + h.c.=\nonumber \\
 &=& S \overline{f_R} \underbrace{\left[ C_L + C_R^{\dagger} \right]}_{\equiv C} f_L + h.c.= \nonumber \\
 &=& S \overline{f_R} C f_L + h.c.,
 \label{eq:scalintsimpler}
\end{eqnarray}
where $C$ is a matrix in flavour space.

The transformation to mass eigenstates leads to $\overline{f_{R,L}}=\overline{f'_{R,L}} U_{R,L}^{\dagger}$ (cf.\ Appendix) and the scalar interaction looks like:
\begin{eqnarray}
\mathcal{L}_{\rm scalar} &=& S \overline{f_R} C f_L + h.c. = S \overline{f'_R} U_R^{\dagger} C U_L f'_L  +h.c.
 \label{eq:scalDtransformed}
\end{eqnarray}
Thereby the condition for complete flavour conservation is:
\begin{equation}
 U_R^{\dagger} \left[ C_L + C_R^{\dagger} \right] U_L \stackrel{!}= {\rm diagonal.}
 \label{eq:scalDcond}
\end{equation}
This condition can be understood as demanding, that the interaction basis is the same as the mass basis. We will refer to such basis identities as {\it alignment}. For the neutral scalars considered here, $C_L$ and $C_R$ can simultaneously be non-zero. To incorporate this interaction into an SM-invariant Lagrangian, the corresponding neutral scalar needs to be a component of an $SU(2)_L$ doublet with hypercharge $1$ or $-1$, that is a copy of the SM Higgs boson or its charge conjugate (with the possible difference of a CP phase -- e.g.\ for the $A$ of a two-Higgs doublet model this phase is just $-1$).

\item[V)] Here, we discuss a neutral intermediate vector boson which can again be defined as real. If it can only couple to left-handed fermions it must be the $T_3=0$ component of an $SU(2)_L$ triplet, i.e.\ a (massive) copy of the SM $W^0$. If it couples to  both left- and right-handed charged leptons it must be an $SU(2)$ singlet, in fact a total singlet under the SM gauge group, that is a (massive) copy of the SM $B^0$. A vector that only couples to right-handed charged leptons is also a total singlet under the SM gauge group, i.e.\ the fact that it does not couple to left-handed charged leptons needs to be explained in the full BSM theory. The interaction Lagrangian is
\begin{eqnarray}
\mathcal{L}_{\rm vector} &=& V_{\mu} \overline{f} \gamma^{\mu} ( C_L \mathcal{P}_L + C_R \mathcal{P}_R )f+h.c. = \nonumber \\
&=& V_{\mu} [\, \overline{f_L} \gamma^{\mu} C_L f_L  + \overline{f_R} \gamma^{\mu} C_R f_R ]+h.c.=\nonumber\\
&=& V_{\mu} [\, \overline{f'_L} \gamma^{\mu} (U_L^\dagger C_L U_L) f'_L  + \overline{f'_R} \gamma^{\mu} (U_R^\dagger C_R U_R) f'_R ]+h.c.,
\label{eq:vecDtransformed}
\end{eqnarray}
where $C_L$ and $C_R$ are necessarily hermitian. To forbid tree-level flavour change, one can demand:
\begin{eqnarray}
 U_L^{\dagger} C_L U_L &\stackrel{!} =& {\rm diagonal,}\nonumber \\
 U_R^{\dagger} C_R U_R &\stackrel{!} =& {\rm diagonal.}
 \label{eq:veclDconds}
\end{eqnarray}
A special case arises if both coefficients, $C_L$ and $C_R$, are proportional to a unit matrix $\mathbbm{1}_F$ in flavour space:
\begin{equation}
 C_L=c_L\cdot \mathbbm{1}_F\ \& \ C_R=c_R\cdot \mathbbm{1}_F.
 \label{eq:flavour_univ}
\end{equation}
This is the {\it flavour universality} condition, as fulfilled e.g.\ for the neutral weak and electromagnetic interactions in the SM. In that case, one gets natural flavour conservation due to the unitarity of the transformation matrices. In case of flavour universality, alignment is automatic, as the identity matrix is the same in all bases. Flavour universality was not an option in the scalar case, as the scalar interaction connects different fermion fields.\\
As the only renormalizable theories of vector bosons are gauge theories~\cite{Itzykson:1980rh}, in general we need to consider these hypothetical, additional vector bosons as gauge particles corresponding to broken generators of some gauge group. Additional vector bosons transforming as an $SU(2)_L$ triplet therefore must be the gauge bosons of gauge group which is broken down to $SU(2)_L$ at some high energy scale. The minimal model in which this is possible uses an $SU(2) \times SU(2)$ gauge group, which is then broken down to its diagonal subgroup $SU(2)_L$. None of the models we discuss introduce such vector bosons, they are however a possible extension of the SM.

\end{enumerate}

We have found that there are only three kinds of neutral particles which can transmit tree-level LFV:

\begin{enumerate}

\item[S) ] A copy of the SM Higgs boson (or its conjugate), with an interaction basis different from the mass basis.

\item[Va) ] A massive copy of the SM photon with flavour non-universal couplings that may or may not discriminate between left- and right-handed particles (which is often called $Z'$).

\item[Vb) ] A copy of the SM $Z$ boson, which is the gauge boson of a gauge group, that is broken down to $SU(2)_L$.

\end{enumerate}

In the following we discuss the Standard Model (SM) and several of its extensions, applying the criteria we have obtained. We do not explicitly mention the cases which are equivalent to the SM case when discussing BSM models. In the following we switch to denoting the involved flavoured fermions by $e$ as most of the results are only applicable to charged leptons. $e=(e,\mu,\tau)^T$ still denotes a vector in flavour space, as mentioned in the Appendix.

%%%%%%%%%%%%%%%%%%%%%%%%%%%%%%%%%%%%%%%%%%%%%%%%%%%%%%%%%%%%%%%%%%%%%%
\subsection{\label{sec:SM} The Standard Model}
%%%%%%%%%%%%%%%%%%%%%%%%%%%%%%%%%%%%%%%%%%%%%%%%%%%%%%%%%%%%%%%%%%%%%%

\begin{center}
\fbox{\underline{Standard Model}: lepton flavour conservation}
\end{center}

As none of the necessary particles is present in the Standard Model of elementary particle physics, we expect no lepton FCNCs at all at tree-level, as we know is the case. To illustrate why this is true and what is exactly ``missing'' in the SM, we give a short discussion.\\
The only neutral scalar in the SM is the usual Higgs boson $H^0$, while for neutral vectors, one has the photon $\gamma$, the $Z^0$ of weak interactions, as well as the gluons $G^a$ from QCD. Hence, the following possibilities remain:

\begin{enumerate}

\item[S)] The only neutral scalar in the SM is the Higgs $H^0$. The SM fermions receive their mass from their coupling to the Higgs field, when this field acquires a non-vanishing vacuum expectation value. Therefore the interaction basis and the mass basis $e'=(e',\mu',\tau')^T$ coincide. Hence the alignment condition is automatically fulfilled, without further restriction of the parameters and the Standard Model has no neutral scalar interaction that could cause flavour non-conservation at tree-level.

\item[Vab)] Under the SM gauge groups all charged leptons in the SM have {\it the same} transformation properties. Their interactions with the SM gauge bosons are therefore flavour universal, which leads directly to the absence of tree-level LFV, as discussed above. We have assumed this flavour universality of the SM interactions in our general discussion and thereby reached the conclusions, that neutral vector bosons with lepton flavour violating couplings must be gauge bosons of a broken symmetry, unrelated to the SM gauge symmetries.\\
We have here implicitly retrieved the original Glashow-Weinberg criteria \cite{Glashow:1976nt}: Criteria 1 and 2 can be understood as demanding flavour universality in the electroweak interactions, while criterion 3 can be reformulated as demanding automatic alignment in the Yukawa sector, which is guaranteed if all fermions receive their mass from one scalar VEV only.

\end{enumerate}

%%%%%%%%%%%%%%%%%%%%%%%%%%%%%%%%%%%%%%%%%%%%%%%%%%%%%%%%%%%%%%%%%%%%%%
\subsection{\label{sec:MultiHiggs} Multi-Higgs Doublet Models}
%%%%%%%%%%%%%%%%%%%%%%%%%%%%%%%%%%%%%%%%%%%%%%%%%%%%%%%%%%%%%%%%%%%%%%

\begin{center}
\fbox{\underline{Multi-Higgs models}: $\forall k: \tilde C_k \stackrel{!}= {\rm diagonal}$ }
\end{center}

As we have seen, there are no lepton FCNCs in the SM. The simplest extensions of the SM are those, where we simply add particles to the SM spectrum. Of the three types of particles which can transmit tree-level lepton FCNCs, a copy of the SM Higgs is the easiest one to add, as it does not require an extension of the SM gauge group. If we add an arbitrary amount of copies of the Higgs boson to the SM, our model is called for obvious reasons a Multi-Higgs Doublet Model. It is simplest to add only one Higgs boson - this is then referred to as a Two-Higgs Doublet Model (THDM) \cite{Gunion:1989we}.

\begin{enumerate}

\item[S)] We can in principle add an arbitrary amount $n$ of {\it additional} Higgs doublets to the SM particle spectrum. These will in general have arbitrary Yukawa couplings to the fermions. The Yukawa Lagrangian for the neutral scalars and charged leptons will therefore be
\begin{equation}
 \mathcal{L} = \sum_{k=1}^{2n+1} H_k \overline{e'_R} C_k e_L' + h.c.,
 \label{eq:Higgsscalint1}
\end{equation}
as we have a total of $(2n+2)$ neutral scalar degrees of freedom (including pseudoscalars), of which one is eaten by the $Z$-boson. One linear combination of all these $H_k$ will have the couplings of the SM Higgs, but this linear combination does not necessarily need to be a mass eigenstate, i.e.\ it will include several different $k$. There is in general no basis where all the $C_k$'s are diagonal, so we consider the above Lagrangian to be written in the mass basis. The condition for absence of tree-level FCNCs is then:
\begin{equation}
 C_k \stackrel{!}= {\rm diagonal},
 \label{eq:Higgsscalint2}
\end{equation}
for all but one $k$ in the mass basis. The last matrix is then automatically diagonal, since we know that one linear combination must be diagonal in the mass basis. This condition leads to well-known constraints such as the Two-Higgs Doublet Models I and II, where an additional $Z_2$ symmetry is imposed, as first discussed in Ref.~\cite{Glashow:1976nt}. Our more general condition for the absence of tree-level FCNCs given above can be rephrased in the following way: We write the Lagrangian in its explicitly $SU(2)_L$ invariant form,
\begin{equation}
 \mathcal{L} = \sum_{k=1}^{n+1} \overline{e_R'} Y_k l_L' \phi_k +h.c.,
 \label{eq:Higgsscalint3}
\end{equation}
where $l_L'$ is the left-handed lepton $SU(2)_L$ doublet, $\phi_k$ is a copy of the SM Higgs doublet, and we are in the mass basis of the charged leptons. Then $C_k$ is diagonal for all $k$ if and only if $Y_k$ is diagonal for all $k$, that is all Yukawa matrices are diagonal in the mass basis. In an arbitrary basis, this means, that given the structure of one Yukawa matrix $Y_k$, all other Yukawa matrices are defined, except for their eigenvalues. So, if we want to forbid tree-level FCNCs, the only new parameter in the Yukawa sector compared to the SM is, for each pair of Higgs boson and fermion, the fraction of the fermion's mass which is generated by the Higgs boson's VEV.\\
In summary, one can say, that the alignment which occurs automatically in the SM is lost in Multi-Higgs models and must be separately postulated to exclude tree-level LFV.

\end{enumerate}

%%%%%%%%%%%%%%%%%%%%%%%%%%%%%%%%%%%%%%%%%%%%%%%%%%%%%%%%%%%%%%%%%%%%%%
\subsection{\label{sec:Z'} $Z'$-models}
%%%%%%%%%%%%%%%%%%%%%%%%%%%%%%%%%%%%%%%%%%%%%%%%%%%%%%%%%%%%%%%%%%%%%%

\begin{center}
\fbox{\underline{$Z'$-models}: $ U_L^\dagger \epsilon'^{(L)} U_L \stackrel{!}= {\rm diagonal}\ \& \ U_R^\dagger \epsilon'^{(R)} U_R \stackrel{!}= {\rm diagonal}$}
\end{center}

$Z'$-type models are also just a very moderate modification of the Standard Model. The general idea is the introduction of an additional flavour non-universal gauge interaction, as opposed to the interactions of the SM, which are flavourblind. The easiest example to consider is the case of one additional gauge boson, corresponding to a new Abelian gauge symmetry $U(1)'$~\cite{Langacker:2000ju}. Of course, this may lead to further complications, such as gauge anomalies and the necessity for additional scalars which break $U(1)'$. However, since we here only concentrate on the lepton flavour violation sector for SM charged leptons, we assume these things to be taken care of.

From our three cases, only Va) is of relevance:

\begin{enumerate}

\item[Va )] One introduces a gauged non-SM symmetry $U(1)'$, under which at least two generations of charged leptons with identical chirality have different charges. This leads to a change in the gauge-covariant derivative, creating an interaction term in the Lagrangian of the form:
\begin{equation}
 \mathcal{L}= - g' \overline{e} \gamma^{\mu} \left[ \epsilon'^{(L)} \mathcal{P}_L +\epsilon'^{(R)} \mathcal{P}_R \right] e Z'_{\mu}.
 \label{eq:Z'vecint1}
\end{equation}
Here, $g'$ is the corresponding coupling constant for the $Z'$-interaction and the charges are absorbed in the coupling matrices. Compared to Eq.~\eqref{eq:vecDtransformed}, we have $V_\mu=Z'_\mu$ and real matrices $C_{L,R}=\epsilon'^{(L,R)}$, adopting the notation of ~\cite{Langacker:2000ju}. The actual vector boson mass eigenstate can in general be a superposition of electroweak and non-SM gauge bosons. Flavour violating couplings can now arise when going to the leptonic mass eigenbasis, if the interactions are flavour non-universal. We start in the eigenbasis of the $Z'$-interaction and hence, the couplings are diagonal. Then, the coupling matrices are given by $\epsilon'^{(L,R)}_{ij}=\epsilon'^{(L,R)}_i \delta_{ij}$, which is flavour non-universal, as long as the $\epsilon'^{(L,R)}_i$'s are not equal. Let $U_L$ and $U_R$ denote the unitary matrices that transform the 3-vectors $e_{L,R}$ in flavour space into their mass eigenstates, $e'_{L,R}=U_{L,R} e_{L,R}$. For the $Z'$-interaction, the Lagrangian then looks like:
\begin{equation}
 \mathcal{L}= -g' \overline{e'} \gamma^{\mu} \left[U_L^\dagger \epsilon'^{(L)} U_L \mathcal{P}_L +U_R^\dagger \epsilon'^{(R)} U_R \mathcal{P}_R \right] e' Z'_{\mu}.
 \label{eq:Z'vecint2}
\end{equation}
The conditions for flavour conservation are:
\begin{equation}
 U_L^\dagger \epsilon'^{(L)} U_L \stackrel{!}= {\rm diagonal}\ \& \ U_R^\dagger \epsilon'^{(R)} U_R \stackrel{!}= {\rm diagonal.}
 \label{eq:Z'vecconds}
\end{equation}

\end{enumerate}
We  can understand these conditions in the following way: If a gauge interaction is no longer flavour universal, the automatic alignment associated with flavour universality is lost, and we need to demand alignment to conserve lepton flavour.

%%%%%%%%%%%%%%%%%%%%%%%%%%%%%%%%%%%%%%%%%%%%%%%%%%%%%%%%%%%%%%%%%%%%%%
\subsection{\label{sec:331} The $331$-model}
%%%%%%%%%%%%%%%%%%%%%%%%%%%%%%%%%%%%%%%%%%%%%%%%%%%%%%%%%%%%%%%%%%%%%%

\begin{center}
\fbox{
\underline{$331$-model}: $U_L^\dagger h_s U_R\stackrel{!}= {\rm diagonal}$}
\end{center}

In the next two sections, we briefly discuss two further extensions of the SM gauge group. Such theories in general lead to additional vector bosons from the extended gauge groups and additional scalars needed to break them down to the SM. The $331$-model is one possible extension of the SM, extending the gauge group to $SU(3)_C \times SU(3)_L \times U(1)_X$, which is then broken to the SM gauge group~\cite{Liu:1993gy}. 

\begin{enumerate}

\item[S)] To break the extended gauge group and give realistic masses to all fermions, three Higgs $SU(3)_L$-triplets ($\Phi$, $\phi$, and $\phi'$) are needed, together with one sextet $H$. Decomposed into SM representations, we are left with three copies of the SM Higgs, of which only two couple to leptons: $\Phi_1^0$, which is part of the $\phi$-triplet and $\Phi_3^0$, which is part of the sextet $H$. In the lepton sector, one is thereby dealing with an effective THDM. In the notation of ~\cite{Liu:1993gy}, the Yukawa interaction for charged leptons is:
\begin{equation}
 \mathcal{L}=-\overline{e_L}\left(\Phi_3^0 h_s+ \Phi_1^0 h_a \right) e_R+h.c.=-\overline{e'_L} U_L^\dagger \left(\Phi_3^0 h_s+ \Phi_1^0 h_a \right) U_R e'_R+h.c.,
 \label{eq:331scalint1}
\end{equation}
where $h_s$ is a symmetric and $h_a$ is an anti-symmetric $3\times3$-matrix in flavour space. As in a THDM, one linear combination of these coupling matrices will always be diagonal in the mass basis, so we only need to demand:
\begin{equation}
 U_L^\dagger h_s U_R\stackrel{!}= {\rm diagonal}.
 \label{eq:331scalint2}
\end{equation}
to prevent tree-level LFV. It should also be noted that, in the $331$-model, flavour changing processes via additional neutral scalars are suppressed due to the smallness of the Yukawa couplings~\cite{Promberger:2007py}.

\item[Va)] To cancel the appearing anomalies, one has to chose one generation of quarks (the third one) to have a transformation behavior different from the other two. The corresponding flavour-changing gauge boson is called $Z'$ and transforms as an SM singlet. No such flavour non-universality is present in the lepton sector however and therefore no tree-level LFV can occur.

\end{enumerate}

%%%%%%%%%%%%%%%%%%%%%%%%%%%%%%%%%%%%%%%%%%%%%%%%%%%%%%%%%%%%%%%%%%%%%%
\subsection{\label{sec:LR} $LR$-symmetric models}
%%%%%%%%%%%%%%%%%%%%%%%%%%%%%%%%%%%%%%%%%%%%%%%%%%%%%%%%%%%%%%%%%%%%%%

\begin{center}
\fbox{\underline{$LR$-symmetric models}: $U_L^\dagger f U_R \stackrel{!}= {\rm diagonal}$}
\end{center}

Another possible extension of the SM gauge group are Left-Right($LR$)-symmetric models \cite{Deshpande:1990ip,Akeroyd:2006bb} with an electroweak gauge group $SU(2)_L \times SU(2)_R \times U(1)_{B-L}$. Here, $R$ stands for ``right'', $B$ is the baryon, and $L$ the lepton number. Then, $SU(2)_R \times U(1)_{B-L}$ is broken down to $U(1)_Y$, which gives the SM. Again, we end up with additional gauge bosons and additional scalars needed to break the enlarged gauge group.

\begin{enumerate}

\item[S)] In order to give masses to the SM fermions, one needs to introduce a Higgs field $\Phi$ transforming as a bi-doublet under $SU(2)_L \times SU(2)_R$. Decomposed into $SU(2)_L$ this results in an adjoint Higgs boson in addition to the SM Higgs. The Yukawa interaction, in the charged lepton mass eigenbasis, is then, employing the notation of \cite{Deshpande:1990ip}:
\begin{equation}
 \mathcal{L}=-\overline{e'_L} U_L^\dagger (f \Phi_2^0 + g \Phi_1^{0*}) U_R e'_R+h.c.,
 \label{eq:LRscalint4}
\end{equation}
which is effectively a two-Higgs doublet model. Comparing with Sec.~\ref{sec:MultiHiggs}, a sufficient condition for the absence of lepton FCNCs in $LR$-symmetric models is:
\begin{equation}
 U_L^\dagger f U_R \stackrel{!}= {\rm diagonal},
 \label{eq:LRscalint5}
\end{equation}
as one linear combination of Yukawa coupling matrices must be diagonal in the mass basis.

\item[Va)] All gauge interactions are in general assumed to be flavour-universal, so we will not encounter tree-level LFVs transmitted by vector bosons.

\end{enumerate}

%%%%%%%%%%%%%%%%%%%%%%%%%%%%%%%%%%%%%%%%%%%%%%%%%%%%%%%%%%%%%%%%%%%%%%
\section{\label{sec:2CC} Tree-level lepton flavour change by doubly charged bosons}
%%%%%%%%%%%%%%%%%%%%%%%%%%%%%%%%%%%%%%%%%%%%%%%%%%%%%%%%%%%%%%%%%%%%%%

For a singly charged scalar or vector, there will be external neutrinos, so we will not consider this case, since we are interested in processes such as $\mu \rightarrow 3e$, where the flavour violation is present for charged leptons. There is only one further way different from FCNCs to mediate such processes already at tree-level, namely by exchanging doubly charged bosons, where again either scalar or vector particles can do the job:

\begin{enumerate}

\item[S)] For a doubly charged scalar, we will have either $C_L=0$ or $C_R=0$ (cf.\ Eq.~\eqref{eq:scalintsimpler}), because otherwise hypercharge would not be conserved. For $C_L \neq 0$, the scalar will be the $T_3=1$ component of an $SU(2)_L$ triplet with hypercharge $Y=2$, i.e.\ of a triplet Higgs. For $C_R \neq 0$ the scalar will be an $SU(2)_L$ singlet with hypercharge $Y=4$. Obviously, a given field cannot have both transformation properties. The Lagrangian reads:
\begin{eqnarray}
 \mathcal{L}_{\rm scalar} &=& S^{++} \overline{(f_L)^\mathcal{C}} C_L f_L + h.c.=\nonumber \\
 &=& S^{++} \overline{(f_L)^\mathcal{C}} C_L f_L + h.c. =\nonumber \\
&=& S^{++} \overline{(f'_L)^\mathcal{C}} U_L^T C_L U_L f'_L + h.c.,
\label{eq:scalDcharged}
\end{eqnarray}
where $L$ can to be replaced with $R$. Note that if the ``left-handed Lagrangian'' given above arises from a triplet Higgs model designed to give mass to the neutrinos, the corresponding doubly charged scalar is in general assumed to be very heavy, giving a further suppression. The condition for absence of tree-level flavour changing diagrams is:
\begin{equation}
U_L^T C_L U_L \stackrel{!}= {\rm diagonal},\ {\rm or}\ U_R^T C_R U_R \stackrel{!}= {\rm diagonal},\ {\rm respectively.}
\label{eq:scalDcharged2}
\end{equation}
We note that in the case of doubly charged scalars, we connect the same fermion field, and therefore we could achieve automatic alignment by demanding flavour universality. For this to work, we would however need $U_{L/R}$ to be real.\\
If these conditions are not fulfilled, one can still fulfill the above condition by demanding the type of alignment defined by Eq.~\eqref{eq:scalDcharged2}, that is alignment for a real $U$.

\item[V)] We can also have doubly charged intermediate vector bosons. These will be $SU(2)_L$ doublet vector bosons with a hypercharge of $Y= +3$. The Lagrangian is:
\begin{eqnarray}
 \mathcal{L}_{\rm vector} &=& V_{\mu}^{++} [\, \overline{(f_R)^\mathcal{C}} \gamma^{\mu} C_L f_L  + \overline{(f_L)^\mathcal{C}} \gamma^{\mu} C_R f_R ] + h.c. =\nonumber \\
 &=& V_{\mu}^{++} [\, \overline{(f'_R)^\mathcal{C}} \gamma^{\mu} U^T_R C_L U_L f'_L  + \overline{(f'_L)^\mathcal{C}} \gamma^{\mu} U^T_L C_R U_R f'_R ] +h.c.
 \label{eq:vecDtransformed2}
\end{eqnarray}
The conditions for the absence of tree-level lepton flavour violation look like:
\begin{eqnarray}
 U^T_R C_L U_L &\stackrel{!}=& {\rm diagonal,}\nonumber \\
 U^T_L C_R U_R &\stackrel{!}=& {\rm diagonal.}
 \label{eq:vecSpecial}
\end{eqnarray}

\end{enumerate}

Flavour universality is of no advantage in this case, so we can only demand the type of alignment defined in the above equation. It is important to note that, apart from leading to tree-level LFV, all the above cases actually produce lepton number violating vertices, or, in other words, the exchange boson has to carry a lepton number. In this case we can give three distinct types of particles, which can mediate doubly charged tree-level LFV:

\begin{enumerate}

\item[Sa) ] An $SU(2)_L$ triplet with hypercharge 2. This particle does not couple to right-handed particles and is equivalent to the triplet Higgs, which is often used for neutrino mass generation.

\item[Sb) ] An $SU(2)_L$ singlet with hypercharge 4. Of the SM fields, this particle can only couple to right-handed charged leptons.

\item[V) ] An $SU(2)_L$ doublet with hypercharge 3. To ensure renormalizability, we must again demand that this vector is a gauge boson. Its gauge group must then contain both, $SU(2)_L$ and $U(1)_Y$, as it is charged under both gauge groups. The smallest gauge group which can contain $SU(2) \times U(1)$ is $SU(3)$. A simple realization is the $331$-model, where the electroweak gauge group is embedded in an $SU(3) \times U(1)$.

\end{enumerate}

After electroweak symmetry breaking, scalar particles of type Sa and Sb can mix.

%%%%%%%%%%%%%%%%%%%%%%%%%%%%%%%%%%%%%%%%%%%%%%%%%%%%%%%%%%%%%%%%%%%%%%
\subsection{\label{sec:Seesaw_2CC} Triplet Higgs Models}
%%%%%%%%%%%%%%%%%%%%%%%%%%%%%%%%%%%%%%%%%%%%%%%%%%%%%%%%%%%%%%%%%%%%%%

\begin{center}
\fbox{\underline{Triplet Higgs}: $U_{\rm PMNS} \stackrel{!}{=} \mathbbm{1}$ (not fulfilled)}
\end{center}

\begin{enumerate}

\item[Sa) ]The simplest models exhibiting tree-level LFV transmitted by doubly charged bosons are again those, where the necessary particles are simply added to the SM. In Triplet Higgs models, an $SU(2)_L$ scalar triplet with hypercharge 2 is added to give Majorana masses to the left-handed neutrinos. To keep the Lagrangian $SU(2)_L$-invariant this scalar also couples to the left-handed charged fermions:
\begin{equation}
 \mathcal{L}= S^{++} \overline{(e_L)^\mathcal{C}} C_L e_L + h.c.,
 \label{eq:See_scal2CC_1}
\end{equation}
which is exactly the Lagrangian of Eq.~\eqref{eq:scalDcharged}. The interaction basis in which $C_L$ is diagonal is that in which the neutrino Majorana mass matrix is diagonal, i.e.\ the neutrino mass basis. To avoid tree-level LFV, $C_L$ should be diagonal in the charged lepton mass basis, i.e.\ the neutrino and charged lepton mass bases must coincide. This would imply that $U_{\rm PMNS}$ is just the unit matrix, which is excluded by experiments. We can therefore say that alignment is experimentally excluded and Triplet Higgs models always induce tree-level LFV, which is however in general strongly suppressed by the large mass of the scalar $SU(2)_L$ triplet.

\end{enumerate}

%%%%%%%%%%%%%%%%%%%%%%%%%%%%%%%%%%%%%%%%%%%%%%%%%%%%%%%%%%%%%%%%%%%%%%
\subsection{\label{sec:331_2CC} The $331$-model}
%%%%%%%%%%%%%%%%%%%%%%%%%%%%%%%%%%%%%%%%%%%%%%%%%%%%%%%%%%%%%%%%%%%%%%

\begin{center}
\fbox{
\begin{tabular}{rl}
\underline{$331$-model}: & $U_L^T h_s U_L \stackrel{!}= {\rm diagonal}$ \& $U_R^T h_s U_R \stackrel{!}= {\rm diagonal}$ (scalars)\\
& $U_R^T U_L \stackrel{!}= {\rm diagonal}$ (vectors)
\end{tabular}
}
\end{center}

In this model, the nearly minimal extension of the SM gauge group, that can generate doubly charged gauge bosons which in turn can mediate LFV, is incorporated. We also encounter doubly charged scalars.

\begin{itemize}

\item[Sab)] In the $331$-model, in general four different doubly charged scalars arise that can couple to leptons and carry a lepton number of $\mp 2$, namely the $T^{\pm \pm}$ and the $\eta^{\pm \pm}$ (note that in the Higgs triplet $\phi'$, another bi-lepton\footnote{A bi-lepton is a particle that carries a lepton number of $\pm 2$.} exists, $\rho^{\pm \pm}$, which however does not couple to leptons and gets its lepton number assignments via terms in the Higgs potential that couple e.g.\ a $\rho^{++}$ and a $\rho^{--}$ with a $T^{++}$ and an $\eta^{--}$, cf.\ Ref.~\cite{Liu:1993gy}). Their couplings to charged leptons look like
\begin{equation}
 \mathcal{L}=-\frac{1}{\sqrt{2}} \overline{e_L} h_s (e_L)^\mathcal{C} T^{++}-\frac{1}{\sqrt{2}} \overline{(e_R)^\mathcal{C}} h_s e_R \eta^{++} + h.c.,
 \label{eq:331_scal2CC_1}
\end{equation}
where $h_s$ was already introduced in Sec.~\ref{sec:331}. Here, the $T^{++}$ is equivalent to the corresponding Sa-particle in a triplet Higgs model, while the $\eta^{++}$ has a hypercharge of $4$ and corresponds to the case Sb. In the mass basis, this gives
\begin{equation}
 \mathcal{L}=-\frac{1}{\sqrt{2}} \overline{e'_L} \underbrace{(U_L^\dagger h_s U_L^*)}_{=(U_L^T h_s U_L)^\dagger} (e'_L)^\mathcal{C} T^{++}-\frac{1}{\sqrt{2}} \overline{(e'_R)^\mathcal{C}} (U_R^T h_s U_R) e'_R \eta^{++} + h.c.,
 \label{eq:331_scal2CC_2}
\end{equation}
from which one can read off the following conditions for flavour conservation:
\begin{eqnarray}
 U_L^T h_s U_L &\stackrel{!}=& {\rm diagonal\ (Sa)},\nonumber\\
 U_R^T h_s U_R &\stackrel{!}=& {\rm diagonal\ (Sb)}.
 \label{eq:331_scal2CC_3}
\end{eqnarray}

\item[V)] Doubly charged massive vector bosons, $Y^{\pm \pm}_\mu$, that get their masses from the $\Phi_Y$ Higgs-doublet, also exist in this model. Their interaction Lagrangian with charged leptons is given by~\cite{Pisano:1991ee}
\begin{equation}
 \mathcal{L}=-\frac{g}{\sqrt{2}} \left[ \overline{(e_R)^\mathcal{C}} \gamma^\mu e_L Y^{++}_\mu + h.c. \right],
 \label{eq:331_vec2CC_1}
\end{equation}
which reads for mass eigenstates
\begin{equation}
 \mathcal{L}=-\frac{g}{\sqrt{2}} \left[Y^{++}_\mu \overline{(e'_R)^\mathcal{C}} \gamma^\mu (U_R^T U_L) e'_L + h.c. \right].
 \label{eq:331_vec2CC_2}
\end{equation}
The condition for the absence of flavour change is
\begin{equation}
 U_R^T U_L \stackrel{!}= {\rm diagonal}.
 \label{eq:331_vec2CC_3}
\end{equation}
Due to the fact, that this gauge interaction couples left- and right-handed charged fermion fields, flavour universality is no longer sufficient for lepton flavour conservation.

\end{itemize}

%%%%%%%%%%%%%%%%%%%%%%%%%%%%%%%%%%%%%%%%%%%%%%%%%%%%%%%%%%%%%%%%%%%%%%
\subsection{\label{sec:LR_2CC} $LR$-symmetric models}
%%%%%%%%%%%%%%%%%%%%%%%%%%%%%%%%%%%%%%%%%%%%%%%%%%%%%%%%%%%%%%%%%%%%%%

\begin{center}
\fbox{\underline{$LR$-models}: $U_{L,R}^T h_{L,R} U_{L,R} \stackrel{!}={\rm diagonal}$}
\end{center}

\begin{itemize}

\item[Sab)] In $LR$-symmetric models, doubly charged Higgses $H_{L,R}^{\pm \pm}$ arise. Their Yukawa couplings are given by~\cite{Akeroyd:2006bb}
\begin{equation}
 \mathcal{L}= H_L^{++} \overline{e^\mathcal{C}}h_L \mathcal{P}_L e + H_R^{++} \overline{e^\mathcal{C}}h_R \mathcal{P}_R e + h.c.= H_L^{++} \overline{(e_L)^\mathcal{C}}h_L e_L + H_R^{++} \overline{(e_R)^\mathcal{C}}h_R e_R + h.c.,
 \label{eq:LR_scal2CC_1}
\end{equation}
Performing the transformations into mass eigenstates and using the formulae from the Appendix, one obtains:
\begin{equation}
 \mathcal{L}=H_L^{++} \overline{(e'_L)^\mathcal{C}}(U_L^T h_L U_L) e'_L + h.c. + (L \leftrightarrow R).
 \label{eq:LR_scal2CC_2}
\end{equation}
Hence, the conditions for the absence of flavour change are:
\begin{equation}
 U_{L,R}^T h_{L,R} U_{L,R} \stackrel{!}={\rm diagonal}.
 \label{eq:LR_scal2CC_3}
\end{equation}
One needs to note here an important difference compared to neutrino mass generation using only a Higgs-triplet: As neutrinos also have Dirac mass terms, due to the presence of right-handed neutrinos, the Yukawa couplings to the Higgs-triplet containing $H_L^{++}$ need not necessarily be diagonal in the neutrino mass basis.

\end{itemize}

%%%%%%%%%%%%%%%%%%%%%%%%%%%%%%%%%%%%%%%%%%%%%%%%%%%%%%%%%%%%%%%%%%%%%%
\section{\label{sec:1-loop} Flavour change at 1-loop level and GIM-suppression}
%%%%%%%%%%%%%%%%%%%%%%%%%%%%%%%%%%%%%%%%%%%%%%%%%%%%%%%%%%%%%%%%%%%%%%

The general form of the amplitude for $\mu\rightarrow e\gamma$ is given in Ref.~\cite{Cheng:1985bj}. One of the results obtained there is, that a chirality flip has to take place during the process, i.e.\ the final electron must have the opposite chirality of the incoming muon. This result is obtained without making any assumptions on the masses of the leptons involved, so that it trivially generalizes to arbitrary flavours and the process $e_i \rightarrow e_k \gamma$. For our purposes the only interesting question is, whether this chirality flip happens on one of the external fermion lines, or arises as net effect of the loop.\footnote{A nice treatment of flavour changing loop diagrams can be found in Ref.~\cite{DeRujula:1977ry}.}\\
In the first case the 1-loop diagram takes the following schematic form (type A: $LL$, type B: $RR$):

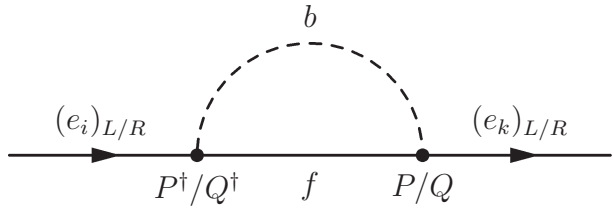
\begin{fmffile}{1-loopAB}
\begin{center}
\begin{fmfgraph*}(80,60)
  \fmfleft{mu}
  \fmf{fermion,label=$(e_i)_{L/R}$,label.side=left}{mu,V1}
  \fmf{plain}{V1,Vf}
  \fmf{plain}{Vf,V2}
  \fmf{fermion,label=$(e_k)_{L/R}$,label.side=left}{V2,e}
  \fmfright{e}
  \fmf{dashes,left,tension=1/3,label=$b$}{V1,V2}
  \fmfdot{V1,V2}
  \fmfv{label=$P^\dagger / Q^\dagger$,label.angle=-90}{V1}
  \fmfv{label=$P / Q$,label.angle=-90}{V2}
  \fmfv{label=$f$,label.angle=-90}{Vf}
 \end{fmfgraph*}
\end{center}
\end{fmffile}
\vspace{-1.2cm}

Note that this is only {\it very} schematic and does not contain several things: First of all, the outgoing photon is missing, that can in general couple either to the internal boson $b$ or to the internal fermion $f$. The diagrams with photons connected to external particles exactly cancel, as discussed in Ref.~\cite{Cheng:1985bj}. This result is again independent of the smallness of the electron mass and hence generalizes to arbitrary flavours. As we are dealing with leptons of the same chirality at both vertices, we also have the same coupling constants (or matrices, in case several distinct particles can appear in the loop) at both vertices. We adopt the general convention that $P$ denotes a coupling matrix involving left-handed leptons, while $Q$ denotes a coupling matrix involving right-handed leptons. We will also in the following refer to diagrams of the above type, i.e.\ with an implicit external helicity flip, as diagrams of type A, if they have external left-handed leptons, and as diagrams of type B if they have external right-handed leptons.

There is now also the possibility of having the chirality flip as net effect of the loop. The 1-loop LFV diagram then takes the following schematic form (type C):

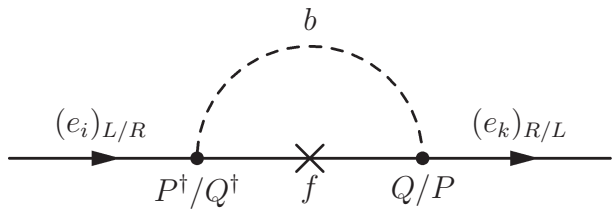
\begin{fmffile}{1-loopC}
\begin{center}
\begin{fmfgraph*}(80,60)
  \fmfleft{mu}
  \fmf{fermion,label=$(e_i)_{L/R}$,label.side=left}{mu,V1}
  \fmf{plain}{V1,Vf}
  \fmf{plain}{Vf,V2}
  \fmf{fermion,label=$(e_k)_{R/L}$,label.side=left}{V2,e}
  \fmfright{e}
  \fmf{dashes,left,tension=1/3,label=$b$}{V1,V2}
  \fmfdot{V1,V2}
  \fmfv{label=$P^\dagger / Q^\dagger$,label.angle=-90}{V1}
  \fmfv{label=$Q / P$,label.angle=-90}{V2}
  \fmfv{decor.shape=cross}{Vf}
  \fmfv{label=$f$,label.angle=-90}{Vf}
 \end{fmfgraph*}
\end{center}
\end{fmffile}
\vspace{-1.2cm}

Again, we have omitted the outgoing photon, as it can couple to either of the internal lines. The chirality flip is now explicitly shown, as a cross on the internal fermion line. We will refer to such diagrams as diagrams of type C. We do not distinguish according to the chirality of the incoming lepton, as in general, if a process where the helicity flips from left to right is possible, the reverse process will be possible as well.

Let us now try to order the conditions under which a flavour change does not occur: First of all, one needs exactly one fermion and one boson in the loop, to ensure Lorentz invariance. In general this means, that we will have one spin-$\frac{1}{2}$-fermion and either a scalar or a vector boson in the loop, as no renormalizable theories for particles with a higher spin are known. Furthermore, SM leptons only carry charge of the gauge groups $SU(2)_L$ and $U(1)_Y$. Hence, this must also be the case for the particle pair in the loop. The internal fermion $f$ may carry e.g.\ a color charge under $SU(3)_C$ (or any other ``exotic'' charge in a theory beyond the SM), as long as this can be compensated by the corresponding internal boson $b$, so that they form a singlet under every gauge group except $SU(2)_L \times U(1)_Y$ ($\times$ possible other groups under which the charged leptons are no singlets). Therefore, another sufficient condition for the absence of flavour change at 1-loop level is
\begin{equation}
 b\otimes f\nsupseteq {\bf 1}\ {\rm (under\ one\ gauge\ group\ except\ } SU(2)_L \times U(1)_Y ).
 \label{eq:1-loop_Cond2}
\end{equation}
These are the obvious criteria for the absence of flavour change. The question remains, what more subtle conditions can be found. Let us consider the three cases we discussed above:

\begin{itemize}
\item[A)] External flip, left-handed charged lepton at both vertices:

In this case we have, at both vertices, a lepton which is the $T_3=-\frac{1}{2}$ component of an $SU(2)_L$ doublet and has hypercharge $Y=-1$. As the photon does not carry away any of these quantum numbers, the tensor product of the internal particles must mimic the transformation properties of the left-handed SM lepton, i.e.\ $b \otimes f\supseteq({\bf 2}_L, Y=-1)$. If no pair of boson and fermion exists with these transformation properties, diagrams of type A are forbidden.

\item[B)] External flip, right-handed charged lepton at both vertices:

Here, the situation is similar to the former case, with the only difference, that the leptons at each vertex are now right-handed. Accordingly, the quantum numbers of the internal particles must (at both vertices) fulfill $b \otimes f \supseteq({\bf 1}_L, Y=-2)$.

\item[C)] Internal flip:

At first sight, this situation seems to be much less straightforward than the other two. At one vertex (the one involving a left-handed external lepton) the boson and the fermion must fulfill the conditions of type A, at the other vertex they must fulfill the conditions of type B. This is naturally only possible after electroweak symmetry breaking. The difference in quantum numbers can only be brought about by a coupling to the Higgs VEV. This can correspond to the mass insertion in the diagram. In that case, the mass insertion serves a double purpose: Inducing the necessary chirality flip and the necessary change in quantum numbers. The chirality flip and the quantum number change can however also be independent of each other, that is if the Higgs VEV couples to the boson line, e.g.\ through a dimension three term. All we definitely need is a coupling to the VEV of the SM Higgs somewhere in the loop.\\
We conclude, that for diagrams of type C to occur, a theory needs a boson and a fermion which fulfill the condition for type A diagrams and another boson-fermion pair that fulfills the condition for type B diagrams. After electroweak symmetry breaking, a superposition of the two fermions gives the mass eigenstate $f$ which appears in the diagram, while a superposition of the two bosons gives the mass eigenstate $b$. Hence, one can say in general, that diagrams of type C are allowed only if both diagrams of type A and of type B are allowed. Note that this condition is necessary, but not sufficient: The mixing of the relevant fermions and bosons is another necessary condition for diagrams of type C to occur.

\end{itemize}

Realizing that these are really the only cases that matter, a third sufficient condition for the absence of flavour change at 1-loop level is
\begin{equation}
 \forall b,f :\ b\otimes f\nsupseteq({\bf 2}_L, Y=-1)\ \&\ b\otimes f\nsupseteq({\bf 1}_L, Y=-2).
 \label{eq:1-loop_Cond3}
\end{equation}
Loop diagrams of the type discussed above even arise in the Standard Model with neutrino masses. They are however strongly suppressed by the GIM-mechanism~\cite{Glashow:1970gm}, which we will generalize in the following.\\
Let $b$ and $f$ be the two particles in the loop. Now let there be $m$ copies of $b$ and $n$ copies of $f$, where copies means that they differ only by their mass. Let $e_i$ denote as before the SM charged leptons. We need to make no assumptions concerning the number of generations, but we assume 3 generations for simplicity. To produce all the above loop diagrams, the Lagrangian must contain the term
\begin{equation}
b_A \overline{(e_{L})_i} P_{iAj} f_j + b_A \overline{(e_{R})_i} Q_{iAj} f_j + h.c.
 \label{eq:1-loop_Cond4}
\end{equation}
For a fixed $A$, the $P_{iAj}$ and $Q_{iAj}$ are in general $3 \times n$-matrices, while for a fixed $j$ they are $3 \times m$-matrices. As they cannot necessarily be diagonalized, since they do not even need to be square matrices, we assume the above term to be written in the mass basis of the SM fermions, the $b_A$ and the $f_j$.\\
This interaction now in general leads to 1-loop flavour-changing diagrams. By a GIM-mechanism, we understand a cancellation of these diagrams, such that the matrix
\begin{equation}
\Gamma_{ik} = \Gamma(e_i \rightarrow e_k \gamma)
\label{eq:1-loop_Cond5}
\end{equation}
is approximately diagonal. If it were exactly diagonal, this would imply, that the matrices $P_{iAj}$ and $Q_{iAj}$ have at most one non-zero entry per column (both for fixed $A$ and fixed $j$). This means explicit conservation of lepton flavour in the interaction, or equivalently that we can assign a specific lepton flavour number to any given boson-fermion pair $b_A$ and $f_j$. Through unitary transformations, any matrices $P_{iAj}$ and $Q_{iAj}$ can be brought to such a form, where they have at most one non-zero entry per column. If they have this form in the respective mass bases of the involved particles, it is another incidence of basis alignment.\\
The GIM-mechanism however means, that we can expand $\Gamma_{ij}$ in some small parameter and the zeroth order coefficient in this expansion is diagonal. This is a slight deviation from our method up until now, as we have so far only considered explicit lepton flavour conservation. However, as this is the mechanism which suppresses LFV in the SM with neutrino masses, and as it relies heavily on the flavour structure of a given model, we find that it is necessary to discuss it here.\\
We give the discussion for left-handed ($Q_{ij}=0$) and fermionic (fixed $A=A_0$, with $b_{A_0}=b$) GIM, where the summation runs over all possible internal fermions $f_j$. This is the case in the SM, with $b$ being the $W$-boson, and the $f_j$ being the light massive neutrinos. It is then straightforward to generalize both to the case of bosonic GIM and to the case of both right-handed and left-handed leptons taking part in the process, i.e.\ $Q_{ij} \neq 0$. The partial decay width for the decay $e_i \rightarrow e_k \gamma$ in the case of left-handed fermionic GIM is \cite{Lavoura:2003xp}:
\begin{equation}
\Gamma_{ik}= \frac{(m_i^2 - m_k^2)^3}{16 \pi m_i^3} (\vert \sum_{j=1}^n P_{ij} P^\dagger_{jk} F(m_i,m_k,m_{f_j},m_b) \vert^2)
\end{equation}
To now obtain the desired result, that is $\Gamma_{ik}$ being approximately diagonal, we need two conditions to be fulfilled. First, we need
\begin{equation}
P P^\dagger \stackrel{!}= {\rm diagonal}
\end{equation}
and second
\begin{equation}
F(m_i, m_k,m_{f_j},m_b) \approx F(m_i,m_k,m_{f_{j'}},m_b)\ \forall j, j' \in \{1,...,n\}\ {\rm and}\ j \neq j'.
\end{equation}
This condition is necessary, so that in a first approximation $F$ can be taken out of the sum and we can use the first condition to diagonalize $\Gamma$. It can be considered as a condition demanding approximate mass degeneracy. What approximate mass degeneracy exactly means is of course ill-defined. The light neutrinos for example are not necessarily approximately degenerate in mass, however, their relative mass differences are small compared to other mass scales in the amplitude, such as the $W$-boson mass, because their absolute mass scale is small. We will not enter further into this discussion, as it is not connected to our main focus, the flavour structure and the particle content of models. It is however important to keep in mind, that, apart from the flavour structure, this approximate mass degeneracy is a necessary condition for the GIM-mechanism to work and thereby for the suppression of 1-loop LFV.\\
Let us also consider the first condition in some more detail. By singular value decomposition, we can write
\begin{equation}
P = U P' V^\dagger
\label{eq:svd}
\end{equation}
where $U$ is a $3 \times 3$ unitary matrix, $V$ is $n \times n$ and also unitary, and $P'$ is a ``diagonal'' $3 \times n$-matrix, that is its only nonzero entries are $P'_{11}$, $P'_{22}$, and $P'_{33}$. Our first condition can then be rewritten as
\begin{equation}
U P' P'^\dagger U^\dagger \stackrel{!}= {\rm diagonal.}
\end{equation}
Our first observation is, that the basis change for the fermions in the loop, given by the matrix $V$, drops out. This is in keeping with the second condition, as for exactly degenerate masses, there would be no uniquely defined mass basis anyhow. Secondly we observe, that $P' P'^\dagger$ is of course diagonal. So, we are again faced with two possibilities: One is that the basis change $U$ defined by Eq.~\eqref{eq:svd} is trivial, that is the mass basis of the charged leptons coincides with the interaction basis, another case of alignment. The other possibility is that $P' P'^\dagger$ is in fact the unit matrix, in which case the above condition is automatically fulfilled - this is the 1-loop equivalent of flavour universality, as the interaction leading to the loop-diagram needs to be just that -- flavour universal.\\
The generalization is then straightforward. In case of the most general interaction, Eq.~\eqref{eq:1-loop_Cond5}, we need to demand
\begin{eqnarray}
P P^\dagger & \stackrel{!}= & {\rm diagonal,} \nonumber \\
Q Q^\dagger & \stackrel{!}= & {\rm diagonal,} \nonumber \\
P Q^\dagger & \stackrel{!}= & {\rm diagonal}
 \label{eq:1-loop_Cond9}
\end{eqnarray}
in the mass basis, where the matrix multiplication is to be understood in such a way, that in each case we either keep $A$ or $j$ fixed.\\
A noteworthy special case is when $f=e$ -- the above condition will then automatically be fulfilled if there is no tree-level LFV (where we assign a separate lepton flavour number to each generation), that is if $P$ and $Q$ are diagonal in the mass basis.\\
The SM with neutrino masses only has GIM-suppressed LFV. In the following discussion, we will not only check for the presence of loop-level LFV, but we will also discuss whether they are GIM-suppressed in the general sense we just have defined.

%%%%%%%%%%%%%%%%%%%%%%%%%%%%%%%%%%%%%%%%%%%%%%%%%%%%%%%%%%%%%%%%%%%%%%
\subsection{\label{sec:SM_loop} The Standard Model}
%%%%%%%%%%%%%%%%%%%%%%%%%%%%%%%%%%%%%%%%%%%%%%%%%%%%%%%%%%%%%%%%%%%%%%

\begin{center}
\fbox{\underline{Standard model}: Diagram A ($f=\nu_{L}$ \& $b=W_\mu^-$) and $P P^\dagger=U_{\rm PMNS}^\dagger U_{\rm PMNS}=\mathbbm{1}$ (GIM)}
\end{center}

The only possibility for a 1-loop level lepton flavour violation like $\mu \rightarrow e\gamma$ in the SM (with massive neutrinos) is diagram A with $b$ being a $W^-$, which also emits the photon, and with $f$ being a neutrino, as will be shown.

The corresponding interaction Lagrangian looks like~\cite{Yao:2006px}
\begin{equation}
 \mathcal{L}=-\frac{e_0}{\sqrt{2} \sin \theta_W} W_\mu^-\, \overline{e_L} \gamma^\mu \nu_L +h.c.= -\frac{e_0}{\sqrt{2} \sin \theta_W} W_\mu^-\, \overline{e'_L} \gamma^\mu U_{\rm PMNS}^\dagger \nu_L'+h.c.
 \label{eq:SM_loop1}
\end{equation}
Now let us go through our criteria: We know that $W^- \sim  ({\bf 3}_L,Y=0)$ and $\nu_L \sim  ({\bf 2}_L,Y=-1)$ with $\nu_L \neq \nu'_L$, while they are singlets under all other gauge groups in the SM (which is just $SU(3)_C$). Hence, also the second sufficient condition for the absence of flavour change is not fulfilled. Now, in $SU(2)$, it holds that ${\bf 3} \otimes {\bf 2}= {\bf 2}\ (\oplus\ {\bf 4})$, so the left-handed neutrino can serve as $f$, since also the hypercharge balance, namely $Y(W^-)-Y(\nu_L)=0-1=-1$ turns out to be the right one. Hence there exists, as expected, lepton flavour violation in the SM, since the mixing matrix $P^\dagger=U_{\rm PMNS}$ is not diagonal. So in the SM, neutrino mixing directly leads to processes like $\mu \rightarrow e\gamma$ at loop-level.

However, the same mixing matrix also leads to GIM-suppression: since $U_{\rm PMNS}$ is unitary, $P P^\dagger=U_{\rm PMNS}^\dagger U_{\rm PMNS}=\mathbbm{1}$ and hence trivially diagonal, which exactly fulfills our condition, Eq.~\eqref{eq:1-loop_Cond9}. This is of course again due to the fact that the weak interaction is flavour universal. As already mentioned, the condition of approximate mass degeneracy is also fulfilled due to the smallness of the absolute neutrino mass scale.

%%%%%%%%%%%%%%%%%%%%%%%%%%%%%%%%%%%%%%%%%%%%%%%%%%%%%%%%%%%%%%%%%%%%%%
\subsection{\label{sec:multiHiggs_loop} Multi Higgs models}
%%%%%%%%%%%%%%%%%%%%%%%%%%%%%%%%%%%%%%%%%%%%%%%%%%%%%%%%%%%%%%%%%%%%%%

\begin{center}
\fbox{
\begin{tabular}{l}
\underline{Multi Higgs models}:\\
A ($f=\nu_R$ or\ $\nu_{heavy}^M$ \& $b=H_k^-$), GIM for $P_k P_k^\dagger \stackrel{!}={\rm diagonal}$;\\
B ($f=\nu_L$ or\ $\nu_{light}^M$ \& $b=H_k^-$), GIM for $Q_k Q_k^\dagger ={\rm diagonal}$; \\
C ($f=\nu_{\rm Dirac}$ \& $b=H_k^-$), GIM for $Q_k U_{\rm PMNS}^\dagger P_k^\dagger \stackrel{!}={\rm diagonal}$
\end{tabular}
}
\end{center}

If we have tree-level LFV in a Multi-Higgs model, we can easily obtain LFV at the 1-loop level by connecting two of the external arms of the tree-level diagram with a mass insertion, giving a diagram of type C. This mass insertion can then also be moved to the two external arms giving diagrams of type A and B. This is a generic statement in models where tree-level LFV is present, so we will not further consider the case of neutral scalars and charged leptons in the loop.\\
We however also can get additional contributions with a charged scalar and a neutrino in the loop. If we do not add right-handed neutrinos to the model, the Lagrangian will contain one relevant interaction, the $SU(2)$-counterterm to the interaction given in Sec.~\ref{sec:MultiHiggs},
\begin{equation}
 \mathcal{L} = \sum_{k=1}^{n} H_k^- \overline{e_R'} Q_k \nu_L + h.c.
 \label{eq:Higgs_loop1}
\end{equation}
Formulated using the general conditions, we have that the additional Higgs bosons transform as $({\bf 2}_L,Y=-1)$ and $\nu_L \sim  ({\bf 2}_L,Y=-1)$, and again they are color singlets, thereby not satisfying the second sufficient condition for the absence of LFV. Taking the product of the representations, we find that $({\bf 2}_L,Y=-1) \otimes ({\bf 2}_L,Y=-1) = ({\bf 1}_L,Y=-2) [\oplus\ ({\bf 3},Y=-2)]$, allowing for diagrams of type B, with $f=\nu_L$ and $b=H_k^-$. As indicated in Eq.~\eqref{eq:Higgs_loop1} we will have $n$ negatively charged scalars: Out of the $(2n+2)$ charged degrees of freedom, half are negative, one of which is eaten by the $W^-$. This implies that in the mass basis no linear combination of $Q_k$ is necessarily diagonal, as that linear combination for neutral scalars corresponds to the eaten scalar in the charged case. As the mass eigenstates of the charged scalars do not necessarily coincide with those of the neutral scalars, the $Q_k$ and the $C_k$ of Sec.~\ref{sec:MultiHiggs} (cf.\ Eq.~\eqref{eq:Higgsscalint1}) are in general not equal. They are however related, since if the original Yukawa coupling matrices $Y_l$ are diagonal for all $l$, then both $C_k$ and $Q_k$ are diagonal for all $k$.\\
This interaction is written in the charged lepton mass basis. This does not coincide with the neutrino mass basis, as we know from the fact that the PMNS-matrix is not diagonal. If we rotate the neutrinos to their mass basis, the interaction reads:
\begin{equation}
 \mathcal{L} = \sum_{k=1}^{n} H_k^- \overline{e_R'} Q_k U_{\rm PMNS}^\dagger \nu_L' + h.c.
 \label{eq:Higgs_loop2}
\end{equation}
So we find that our coupling matrix $Q_k U_{\rm PMNS}^\dagger$ is not diagonal, that is diagrams of type B are allowed, even if tree-level LFV is forbidden. The condition for GIM-suppression then reads:
\begin{equation}
 Q_k U_{\rm PMNS}^\dagger U_{\rm PMNS} Q_k^\dagger = Q_k Q_k^\dagger \stackrel{!}= {\rm diagonal.}
 \label{eq:Higgs_loop3}
\end{equation}
This means that, if tree-level LFV is forbidden, and thereby $Q_k$ is diagonal, these processes will always be GIM-suppressed. As in the SM this is also due to the fact, that the absolute mass scale of the light neutrinos is small compared to the mass of scalars.\\
If we add three right-handed neutrinos to the model, there are two possibilities: One can either write down a Majorana mass term for the right-handed neutrinos and apply the Type I seesaw mechanism or one can consider neutrinos as Dirac particles.\\
In the first case, the mass eigenstates will be Majorana particles, a superposition of left-handed and right-handed neutrinos. We will write these as $\nu^M_{light}$ for the predominantly left-handed light neutrinos and as $\nu^M_{heavy}$ for the predominantly right-handed heavy neutrinos. As we have so far always assumed a unitary PMNS-matrix, i.e.\ a ``perfect'' seesaw, we will assume that the the light neutrinos are purely left-handed and the heavy ones are purely right-handed.\footnote{Limits on the non-unitarity of the PMNS-matrix are considered in Ref.~\cite{Antusch:2006vw}} We then still have the interaction of Eq.~\eqref{eq:Higgs_loop2} with $\nu_L'$ replaced by $\nu^M_{light}$ and the same conditions for LFV and GIM-suppression in diagrams of type B. The right handed or heavy neutrinos transform as total singlets under the SM gauge groups, so that the product of their representation with that of the Higgs bosons is $({\bf 2}_L,Y=-1)$, allowing for diagrams of type A, with $f=\nu^M_{heavy}$ and $b=H_k^-$.\\
The corresponding couplings are then the Yukawa couplings which give the neutrinos their Dirac mass:
\begin{equation}
 \mathcal{L} = \sum_{k=1}^n H_k^- \overline{e_L'} P_k \nu^M_{heavy} + h.c.
 \label{eq:Higgs_loop4}
\end{equation}
As the matrices $P_k$ play no role for tree-level LFV, we make no further assumptions concerning their form. However one needs to pay close attention in which basis the above interaction is written: We have chosen the basis in which the charged lepton and the right-handed neutrino Majorana matrices are diagonal. If we had written the interaction in another basis, one would here also have to introduce a PMNS-type matrix, as was the case for the light neutrinos. LFV-processes will then occur if $P_k$ is not diagonal and the condition for GIM-suppression is then simply
\begin{equation}
 P_k P_k^\dagger \stackrel{!}= {\rm diagonal.}
 \label{eq:Higgs_loop5}
\end{equation}
This GIM-suppression of course demands, that the heavy neutrinos are approximately degenerate in mass. Such processes however are strongly suppressed anyway, as the heavy neutrinos decouple in the seesaw limit. Even though diagrams of type A and B are allowed, no diagrams of type C can be generated for Majorana neutrinos, as the necessary condition that the fermions of diagrams A and B mix after electroweak symmetry breaking is not fulfilled in the seesaw limit.\\
Things are different for the case of Dirac neutrinos. One again has the interactions of Eq.~\eqref{eq:Higgs_loop2} and of Eq.~\eqref{eq:Higgs_loop4}, where $\nu^M_{heavy}$ must be replaced by $\nu_R'$, the right-handed neutrinos in the neutrino mass basis. This means diagrams of type A and type B can occur under the same conditions as above. As left- and right-handed neutrinos mix in this case to form a Dirac fermion after electroweak symmetry breaking, diagrams of type C are now also possible, if either $Q_k U_{\rm PMNS}^\dagger$ or $P_k$ is not diagonal. As we assume no tree-level LFV, $Q_k U_{\rm PMNS}^\dagger$ is automatically non-diagonal and such processes can occur. The condition for GIM-suppression is then
\begin{equation}
 Q_k U_{\rm PMNS}^\dagger P_k^\dagger \stackrel{!}= {\rm diagonal.}
 \label{eq:Higgs_loop6}
\end{equation}
We reach the conclusion, that if tree-level LFV is forbidden in a Multi-Higgs model, then 1-loop LFV including left-handed neutrinos will always be GIM-suppressed. Observable LFV therefore necessitates the introduction of right-handed neutrinos. As these will approximately decouple in the Majorana case, only Dirac neutrinos lead to observable 1-loop LFV-processes. A GIM-suppression of such processes could then only be brought about by demanding the alignment conditions of Eqs.~\eqref{eq:Higgs_loop5} and \eqref{eq:Higgs_loop6}.

%%%%%%%%%%%%%%%%%%%%%%%%%%%%%%%%%%%%%%%%%%%%%%%%%%%%%%%%%%%%%%%%%%%%%%
\subsection{\label{sec:UED_loop} Universal Extra Dimensions}
%%%%%%%%%%%%%%%%%%%%%%%%%%%%%%%%%%%%%%%%%%%%%%%%%%%%%%%%%%%%%%%%%%%%%%

\begin{center}
\fbox{
\begin{tabular}{l}
\underline{Universal Extra Dimensions}:\\
A ($f=\nu_{L(n)}$ \& $b=W_{\mu (n)}^-$), as for SM (GIM); \\
A ($f=\nu_{R(n)}$ \& $b=a_{(n)}^-$), where $P=U_{\rm PMNS}^\dagger c_R$ and $P P^\dagger \propto \mathbbm{1}$ (GIM);\\
B ($f=\nu_{L(n)}$ \& $b=a_{(n)}^-$), where $Q=U_{\rm PMNS}^\dagger c_L$ and $Q Q^\dagger \propto {\rm diag}(m_e^2,m_\mu^2,m_\tau^2)$ (GIM);\\
C ($f=\nu_{R/L(n)}$ \& $b=a_{(n)}^{-}$), $P Q^\dagger \propto {\rm diag}(m_e,m_\mu,m_\tau)$ (GIM)
\end{tabular}
}
\end{center}

A different type of models where lepton flavour violation can occur are theories with extra spatial dimensions. There is a huge variety of them - we will only be considering the  ACD-model~\cite{Appelquist:2000nn}, which is also often called Universal Extra Dimensions (UEDs). A key feature of this model is, that the particles of the SM propagate in all 5 dimensions, where the 5th dimension is compactified.\\
We adopt the notation of Ref.~\cite{Buras:2002ej}. In this model, there are then two types of particles that can play the role of the boson $b$. First of all we have the vector bosons $W_{(n)}^-$, where $n$ denotes the KK-number. These KK-modes of the $W$-boson transform in the same way as the zero mode, which is just the SM $W$, under all SM gauge groups. Hence, we know from Sec.~\ref{sec:SM_loop} that a particle transforming as a left-handed neutrino can here be used as the fermion $f$. UEDs lead to an additional symmetry which needs to be conserved, the conservation of the KK-number $n$. To ensure that the particles in the loop form a total singlet under all non-SM symmetries, we need to demand that the neutrino-like particle in the loop has the same KK-number as does the boson. Therefore the only particle that can here play the role of $f$ is the $n$-th KK-mode of the neutrino, $\nu_{(n)}$. Otherwise, nothing changes compared to the SM with massive neutrinos: Diagrams of type A will be allowed and GIM-suppression will always occur due to the unitarity of $U_{\rm PMNS}$. The $n$-th KK-mode of a given neutrino $\nu_i$ will have mass $m^2_{(n)}=m^2_i+\frac{n^2}{R^2}$, where $m_i$ is the zero-mode mass of the neutrino and $R$ is the compactification radius of the extra dimension. Hence, the mass degeneracy is even more explicit here, as the mass differences of neutrino KK-modes are small compared to their mass, which is approximately $\frac{n}{R}$.\\
UEDs also lead to scalars that can take the part of $b$: The higher KK-modes of the charged and pseudoscalar Higgs fields are not entirely eaten by the corresponding vector bosons, they also mix with the 5th component of those vector bosons to form physical scalars, both charged ($a_{(n)}^-$) and neutral ($a_{(n)}^0$). As these scalars transform as the SM Higgs, they can form a loop with particles transforming as neutrinos or as charged leptons, as discussed in Sec.~\ref{sec:multiHiggs_loop}. Again, we need to observe conservation of KK-number, so $f$ can only be $\nu_{(n)}$ or the $n$-th KK-mode of the charged lepton, $e_{(n)}$, respectively. So, we will have the same types of diagrams as in a Multi-Higgs model, with the particles in the loop replaced by their higher KK-modes. As opposed to a Multi-Higgs model, the additional scalars can be considered as excitations of the same particle, and therefore all couple in the same way. They will however couple differently from the SM Higgs, as they also have a gauge boson contribution. All gauge interactions however remain flavour universal, so we find, for the coupling of left-handed charged leptons to $e_{(n)}$ and $a_{(n)}^0$:
\begin{equation}
P \propto (Y_e + \mbox{flavour universal contributions}).
\end{equation}
For the coupling of the right-handed charged leptons we have no further complications from gauge interactions and the coupling matrices $Q$ will just be proportional to the regular charged lepton Yukawa couplings $Y_e$. One can then see, that all coupling matrices are diagonal in the charged lepton mass basis, which is also the mass basis for the KK-modes $e_{(n)}$, and we therefore have no LFV for diagrams with $e_{(n)}$ in the loop.\\
For neutrino KK-modes in the loop, we obtain the Lagrangian~\cite{Bigi:2006vc}
\begin{eqnarray}
 \mathcal{L} & = & -\frac{g_2 n}{\sqrt{2} M_{W(n)}} \left[ \overline{\nu_{R(n)}} c_R e_L' + \overline{\nu_{L(n)}} c_L e_R' \right] a_{(n)}^-+h.c. \\
& = & -\frac{g_2 n}{\sqrt{2} M_{W(n)}} \left[\, \overline{\nu'_{R(n)}} U_{\rm PMNS}^{\dagger} c_R e_L' + \overline{\nu'_{L(n)}} U_{\rm PMNS}^{\dagger} c_L e_R' \right] a_{(n)}^- +h.c.,
 \label{eq:UED_loop1}
\end{eqnarray}
with $c_L= {\rm diag}(m_e,m_\mu,m_\tau)$ and $c_R= M_W \cdot \mathbbm{1}$. Note that in principal there can be a correction to $c_R$ coming from the neutrino Yukawa coupling matrix, but we will assume neutrinos to be purely Dirac. In that case their masses are negligible compared to $M_W$ and can be ignored in Eq.~\eqref{eq:UED_loop1}. For comments on different methods of neutrinos mass generation and their effect on LFV, see Sec.~\ref{sec:multiHiggs_loop}. The right-handed neutrinos in Eq.~\eqref{eq:UED_loop1} are KK-modes of the left-handed neutrino and so arise independently of the origin of neutrino mass. We find that the relevant coupling matrices $P$ and $Q$ are both the product of a flavour-diagonal matrix and the non-diagonal, unitary $U_{\rm PMNS}$. We therefore can construct LFV diagrams of all types; all such processes will however be GIM-suppressed, as the mass degeneracy of the $\nu_{(n)}$ is again explicit. See Ref.~\cite{Bigi:2006vc} for a discussion of the effect of summing over a large number of GIM-suppressed amplitudes.

%%%%%%%%%%%%%%%%%%%%%%%%%%%%%%%%%%%%%%%%%%%%%%%%%%%%%%%%%%%%%%%%%%%%%%
\subsection{\label{sec:MSSM_loop} The Minimal Supersymmetric Standard Model}
%%%%%%%%%%%%%%%%%%%%%%%%%%%%%%%%%%%%%%%%%%%%%%%%%%%%%%%%%%%%%%%%%%%%%%

\begin{center}
\fbox{
\begin{tabular}{l}
\underline{MSSM$+\nu_R$}:\\
 A ($f=(\tilde \chi_{A,R}^{-/0})'$ \& $b=\tilde \nu'/ \tilde e'$), where $P=(C/N)_{A}^{R(l)}$, and $P P^\dagger={\rm diagonal}$ (GIM);\\
 B ($f=(\tilde \chi_{A,L}^{-/0})'$ \& $b=\tilde \nu'/ \tilde e'$), where $Q=(C/N)_{A}^{L(l)}$, and $Q Q^\dagger = {\rm diagonal}$ (GIM);\\
 C ($f=(\tilde \chi_{A}^{-/0})'$ \& $b=\tilde \nu'/ \tilde e'$), where $P Q^\dagger={\rm diagonal}$ (GIM)
\end{tabular}
}
\end{center}

The MSSM itself can only lead to 1-loop LFV diagrams, since all tree-level vertices are forbidden due to $R$-parity conservation. The discussion is somewhat similar to that of Sec~\ref{sec:UED_loop}, as we again take the diagrams of the SM and Multi-Higgs models, and replace the particles in the loop by other particles which transform in the same way under the SM gauge groups, thereby delegating a large part of the discussion concerning the general LFV-conditions to Secs.~\ref{sec:SM_loop} and \ref{sec:multiHiggs_loop}. In the case of the MSSM, the particles in the loop will be replaced by their superpartners, thereby also ensuring that there is always one boson and one fermion in the loop.\\
We begin by considering the supersymmetric analogon of the LFV diagrams with neutrinos in the loop. The neutrinos will be replaced by sneutrinos, which are then the bosons in the loop, $b=\tilde\nu$. In the MSSM, the LFV diagrams with a $W$ in the loop (Sec.~\ref{sec:SM_loop}) and with a charged Higgs scalar in the loop (Sec.~\ref{sec:multiHiggs_loop} - they arise as the MSSM is a THDM) are both replaced by diagrams with charginos, which are then the fermions in the loop, $f=\tilde \chi^-_A$ ($A=1,2$). This is because the two $\tilde \chi^-_L$'s are superpositions of the gaugino $\tilde W^-_L$ (superpartner of the $W$) and the Higgsino $\tilde H^-_{uL}$ (superpartner of one Higgs boson), and conversely the two $\tilde \chi^-_R$'s are superpositions of $\tilde W^-_R$ and $\tilde H^-_{dR}$. The sneutrinos will be massive, even if the neutrinos are not, due to soft SUSY breaking, so we do not need to worry about the origin of neutrino mass. The sneutrino mass basis need not coincide with that of the charged leptons and we expect LFV to occur. The interaction Lagrangian is\footnote{For more details, see Refs.~\cite{Hisano:1995cp} and \cite{Chacko:2001xd}.}
\begin{equation}
 \mathcal{L}_{\rm chargino}= \sum_{A=1}^2
 \overline{e'_{L}} C_{A}^{R(l)} (\tilde \chi_{A,R}^-)' \tilde \nu'+
 \overline{e'_{R}} C_{A}^{L(l)} (\tilde \chi_{A,L}^-)' \tilde \nu'+h.c.
 \label{eq:MSSM_loop5}
\end{equation}
Here, all fields (also the bosonic ones) are written as mass eigenstates, and $C_{A}^{R/L(l)}$ denotes the coupling matrix of the right- and left-handed chargino $(\tilde \chi_{A,R/L}^-)'$, respectively, to the charged leptons $e$ and the sneutrino mass eigenstates $\tilde \nu'$. The $C_{A}^{R/L(l)}$ are thereby $3 \times 3$-matrices. These matrices contain all rotations to mass eigenstates, for the left- and right-handed charginos as well as for the sneutrinos. For diagram A, one then needs $f=(\tilde \chi_{A,R}^-)'$ and $P=C_{A}^{R(l)}$ has to be non-diagonal. Diagram B is possible with $f=(\tilde \chi_{A,L}^-)'$, and $Q=C_{A}^{L(l)}$ non-diagonal, and diagram C with $f$ flipping from $(\tilde \chi_{A,L}^-)'$ to $(\tilde \chi_{A,R}^-)'$, or vice versa, with the same $P$ and $Q$ as before. If all $C_{A}^{R/L(l)}$ turn out to be diagonal there is no LFV at 1-loop level -- this is of course a case of alignment, a term which actually appears to originate from the supersymmetric case \cite{Nir:1993mx}. If the $C_{A}^{R/L(l)}$ are not diagonal, there is still the possibility of GIM-suppression, for which the conditions are
\begin{eqnarray}
 C_{A}^{R(l)} C_{A}^{R(l)\dagger} &\stackrel{!}=& {\rm diagonal\ (A),}\\
 C_{A}^{L(l)} C_{A}^{L(l)\dagger} &\stackrel{!}=& {\rm diagonal\ (B),}\\
 C_{A}^{R(l)} C_{A}^{L(l)\dagger} &\stackrel{!}=& {\rm diagonal\ (C).}
 \label{eq:MSSM_loop6}
\end{eqnarray}
These conditions are in fact always fulfilled: Since the chargino is a superposition of Higgsino and wino, we need to invoke the natural alignment of mass and Yukawa interaction basis as well as the flavour universality of the weak interaction. LFV only arises due to the non-trivial, unitary transformations to mass eigenstates. The critical question is therefore the approximate mass degeneracy, which can be achieved by giving approximately universal soft masses to the sneutrinos. Their mass differences, corresponding to the mass differences of the neutrinos, then become negligible. This is commonly referred to as the Super-GIM mechanism \cite{Dimopoulos:1981zb} and is in fact covered by the generalized GIM-mechanism.\\
In general, LFV diagrams with charged leptons in the loop are only allowed if tree-level LFV is also allowed (cf.\ Sec.~\ref{sec:multiHiggs_loop}). As the KK-modes of the charged leptons necessarily have the same mass basis as the charged leptons themselves, the ``partner'' diagrams for UEDs also did not lead to LFV (cf. Sec.~\ref{sec:UED_loop}). Things are different in the MSSM, as the superpartners of the charged leptons, the charged sleptons $\tilde e$, do not necessarily have the same mass basis, because their mass also arises from soft SUSY breaking terms. Basis alignment can be achieved by imposing conditions on the soft SUSY breaking terms, such as the popular mSUGRA boundary conditions, but in general one can construct diagrams with charged sleptons in the loop taking the role of $b$. The part of $f$ is then taken by a superposition of the superpartners of the neutral electroweak gauge bosons and the neutral Higgs bosons. The mass eigenstates are the neutralinos $\tilde \chi^0_A$ ($A=1,...,4$), where the $\chi^0_A$ is a superposition of the bino $\tilde B$, the neutral wino $\tilde W^0$, and the two neutral Higgsinos $\tilde H^0_u$ and $\tilde H^0_d$). The corresponding interaction Lagrangian is
\begin{equation}
 \mathcal{L}_{\rm neutralino}= \sum_{A=1}^4
 \overline{e'_{L}} N_{A}^{R(l)} (\tilde \chi_{A,R}^0)' \tilde e'+
 \overline{e'_{R}} N_{A}^{L(l)} (\tilde \chi_{A,L}^0)' \tilde e'+h.c.,
 \label{eq:MSSM_loop7}
\end{equation}
where the matrices $N_{A}^{R/L(l)}$ now contain the rotations of $(\tilde B, \tilde W^0, \tilde H^0_u, \tilde H^0_d)^T$ to mass eigenstates $((\tilde \chi_{1}^0)',...,(\tilde \chi_{4}^0)')^T$ for both cases, $R$ and $L$, and the rotations of the charged sleptons to mass eigenstates, too. The cases that can appear here are completely analogous to the ones for charginos, just with $(\tilde \chi^-)' \rightarrow (\tilde \chi^0)'$, $\tilde \nu' \rightarrow \tilde e'$, and $C \rightarrow N$ for the mixing matrices. The only difference is that there exist four different neutralinos compared to only two negatively charged charginos and six charged sleptons, making the $N_{A}^{R/L(l)}$ $6 \times 3$-matrices. Again the flavour condition for GIM-suppression is automatically fulfilled, while the mass degeneracy can be achieved by approximately universal soft masses.

%%%%%%%%%%%%%%%%%%%%%%%%%%%%%%%%%%%%%%%%%%%%%%%%%%%%%%%%%%%%%%%%%%%%%%
\section{\label{sec:Summary} Summary and Conclusions}
%%%%%%%%%%%%%%%%%%%%%%%%%%%%%%%%%%%%%%%%%%%%%%%%%%%%%%%%%%%%%%%%%%%%%%

In this paper, we have given general criteria a theory has to fulfill in order to avoid or to allow for LFV-processes. We have found that one can indeed give very simple conditions that are sufficient for such statements, at least for tree-level and 1-loop diagrams. These conditions only refer to the particle content and the flavour structure of the couplings of a given model.

As first possibility, we have considered the cases of neutral (Sec.~\ref{sec:FCNCs}) and doubly charged exchange bosons (Sec.~\ref{sec:2CC}), where the latter ones do not occur in the SM. We have distinguished  between scalar and vector bosons, and have identified the SM transformation properties of all particles that could mediate the respective process. We discussed how tree-level LFV-processes can be prevented, even in the presence of such particles, by demanding alignment or flavour universality in the flavour structure of the couplings. We have attempted to give the conditions in a concise and easily applicable way. To test their applicability, we then applied our criteria to several models.

In Sec.~\ref{sec:1-loop} we have investigated how LFV-processes can occur at 1-loop level, again first determining the necessary particle content. We also studied the general cases for GIM-suppressed amplitudes, using a generalization of the GIM-mechanism. The ideas of alignment and flavour universality were also generalized to the 1-loop case. We found that, even if the loop-processes look much more complicated than their tree-level analogues, one can still narrow down the necessary ingredients for a flavour change to very simple requirements, for the occurrence of the processes themselves as well as for a possible GIM-suppression. Again, we have investigated the situation for some exemplary models in order to clarify our criteria and to prove their applicability.

A complete summary of our results can be found in the summary table on the next page. Notations and conventions we have used are listed in the Appendix.

\clearpage
\begin{sideways}
\begin{minipage}{\textheight}
\begin{tabular}{|c|c|c|c|c|}
\hline
Model & Conditions for the  & Conditions for the absence      & Conditions for the & 1-loop flavour change\\
      & absence of tree-level & of tree-level FC by doubly    & absence of 1-loop  & \& GIM-suppression \\
      & FCNCs (S/V) & charged bosons (Sab/V) & flavour change     & \\ \hline \hline
SM             & S) aut.\ Align & N/A & A: ($\nu_{L}$, $W_\mu^-$), Align excl. & A: aut.\ GIM\\
               & Vab) FU    & & & \\ \hline
multi Higgs    & S) Align & N/A & A: ($\nu^M_{heavy} / \nu_R$, $H_k^-$), Align & A: Align, mass deg.\ for $\nu^M_{heavy}$ \\
               & & & B: ($\nu_L/\nu_{light}^M$, $H_k^-$), Align excl. & B: aut.\ GIM \\
               & & & C: ($\nu_{\rm Dirac}$, $H_k^-$), Align excl. & C: Align \\
               \hline
$Z'$           & Va) Align & N/A & N/I & N/I \\
               \hline
Triplet Higgs  & N/A & Sa) Align excl. & N/I & N/I \\
               \hline
$331$          & S) Align & Sab) Align & N/I & N/I \\
               & Va) FU & V) Align & & \\
               \hline
$LR$           & S) Align & Sab) Align & N/I & N/I \\
               & Va) FU & & & \\ \hline
UEDs           & N/A & N/A & A: ($\nu_{L(n)}$, $W_{\mu(n)}^-$), Align excl. & A, B, C: aut. GIM  \\
               & & & A: ($\nu_{R(n)}$, $a_{(n)}^-$), Align excl. & \\
               & & & B: ($\nu_{L(n)}$, $a_{(n)}^-$), Align excl. & \\
               & & & C: ($\nu_{R/L(n)}$, $a_{(n)}^-$), Align excl.  &\\                        \hline
MSSM $+\nu_R$  & N/A without $R$- & N/A & A: ($(\tilde \chi_{A,R}^{-/0}$)', $\tilde \nu'/ \tilde e'$), Align                 & A, B, C:  \\
               & parity & & B: ($(\tilde \chi_{A,L}^{-/0})'$, $\tilde \nu'/ \tilde e'$), Align & aut.\ Align, mass deg. \\
               & & & C: ($(\tilde \chi_{A}^{-/0})'$, $\tilde \nu'/ \tilde e'$), Align &  \\ \hline
\end{tabular}
Summary table of our results. (N/A: not applicable; N/I: not investigated, due to presence at tree-level; aut.\ Align: Alignment is automatic in this model; Align excl.: Alignment is excluded phenomenologically; Align: Alignment needs to be imposed; FU: flavour universality; aut.\ GIM: all GIM-conditions are automatically fulfilled; mass deg.: Mass degeneracy needs to be imposed)\\
\end{minipage}

\end{sideways}

\clearpage

%%%%%%%%%%%%%%%%%%%%%%%%%%%%%%%%%%%%%%%%%%%%%%%%%%%%%%%%%%%%%%%%%%%%%%
\section*{Acknowledgements}
%%%%%%%%%%%%%%%%%%%%%%%%%%%%%%%%%%%%%%%%%%%%%%%%%%%%%%%%%%%%%%%%%%%%%%

We would like to thank T.~Ota and M.~Lindner for useful comments as well as C.~Hagedorn for organizing our seminar on LFV. This work has been supported by the DFG-Sonderforschungsbereich Transregio 27 ``Neutrinos and beyond -- Weakly interacting particles in Physics, Astrophysics and Cosmology''. AB acknowledges support from the Studienstiftung des Deutschen Volkes.

%%%%%%%%%%%%%%%%%%%%%%%%%%%%%%%%%%%%%%%%%%%%%%%%%%%%%%%%%%%%%%%%%%%%%%
\section*{\label{sec:conventions} Appendix: Notations \& Conventions}
%%%%%%%%%%%%%%%%%%%%%%%%%%%%%%%%%%%%%%%%%%%%%%%%%%%%%%%%%%%%%%%%%%%%%%

\begin{itemize}

\item $\mathcal{P}_{L,R}\equiv \frac{1\mp \gamma_5}{2}$: left- and right-handed projection operator (properties: $\mathcal{P}_{L/R}^2=\mathcal{P}_{L/R}$, $\mathcal{P}_{L}+\mathcal{P}_{R}=1$, $\mathcal{P}_{R}\mathcal{P}_{L}=\mathcal{P}_{L}\mathcal{P}_{R}=0$)

\item Charge conjugation: $\Psi^{\mathcal{C}}\equiv C (\overline{\Psi})^T$ with $C=i\gamma^2 \gamma^0$ (properties: $C\gamma_{\mu}^T C^{-1}=-\gamma_{\mu}$, $C^{-1}=-C=C^T=C^\dagger$)

\item Relations with projection operators (using $\gamma_5^\dagger=\gamma_5^T=\gamma_5^*=\gamma_5$ and $\{\gamma^\mu,\gamma_5\}=0$ which leads to $\gamma^\mu \mathcal{P}_{L,R}=\mathcal{P}_{R,L} \gamma^\mu$):
 \begin{eqnarray}
  \mathcal{P}_{L,R}\Psi&=&\Psi_{L,R}\nonumber\\
  \mathcal{P}_{L,R}\Psi^\mathcal{C}&=&(\Psi_{R,L})^\mathcal{C} \nonumber\\
  \overline{\Psi}\mathcal{P}_{L,R}&=&\overline{\Psi_{R,L}}\nonumber\\
  \overline{\Psi^\mathcal{C}}\mathcal{P}_{L,R}&=&\overline{(\Psi_{L,R})^\mathcal{C}}\nonumber
  \label{eq:projections}
 \end{eqnarray}

\item Dirac mass terms:\\
 A mass term for a general vector $f=(f_1,f_2,...,f_N)^T$ of Dirac fermions in an $N$-dimensional flavour space is given by
\begin{equation}
 \mathcal{L}_{\rm Dirac}=-\overline{f_R} M_D f_L-\overline{f_L} M_D^\dagger f_R,
\end{equation}
where $M_D \in \mathbb{C}^{N\times N}$ is an arbitrary matrix in the $N\times N$ flavour space. Hence, it can be diagonalized by a bi-unitary transformation leading to
\begin{equation}
 D_D={\rm diag}(m_1,m_2,...,m_N)=U_L M_D^\dagger U_R^\dagger=U_R M_D U_L^\dagger,\ {\rm with}\ m_i>0,
\end{equation}
where $U_{L,R}^{\dagger}=U_{L,R}^{-1}$ \& $U_L M_D^\dagger M_D U_L^\dagger=U_R M_D M_D^\dagger U_R^\dagger=D_D^2$.

Hence the transformation of $f$ (which in this work will -- if not differently stated -- be an eigenstate of the respective interaction) to the mass eigenstate $f'$ is given by
\begin{equation}
 f_R = U_R f'_R\ \& \ f_L=U_L f'_L.
\end{equation}

\item Transformations of the eigenstates $\Psi$ of an interaction into the mass eigenstates $\Psi'$ (viewing $\Psi$ as vector in flavour space and keeping in mind that e.g.\ $\gamma$-matrices that act on spinors and hence on the {\it components} of $\Psi$ must commute with a matrix $U$ in flavour space, since for such a matrix $U$ they only look like scalars):
 \begin{eqnarray}
  \Psi&=&U\Psi'\nonumber\\
  \Psi^\mathcal{C}&=&U^* {\Psi'}^\mathcal{C}\nonumber\\
  \overline{\Psi}&=&\overline{\Psi'}U^\dagger\nonumber\\
  \overline{\Psi^\mathcal{C}}&=&\overline{{\Psi'}^\mathcal{C}}U^T\nonumber
  \label{eq:mass_transformations}
 \end{eqnarray}

\item Relation between vectors in flavour space, $SU(2)$-doublets, and spinors: $e=(e,\mu,\tau)^T$ (components are 4-spinors) and $l=(\nu,e)^T$ (components are vectors in flavour space, whose components are 4-spinors)

\item Convention for the PMNS-matrix: $\nu=U_{\rm PMNS}^\dagger \nu'$ (the complete mixing happens in the neutrino sector, as usual)

\item Charged lepton quantum numbers ($SU(2)_L$-representation, weak isospin $T_3$, hypercharge $Y$, electric charge $Q$ as obtained by $Q=T_3+\frac{Y}{2}$, and $\gamma_5$-eigenvalue):
\begin{center}
\begin{tabular}{|c||c|c|c|c|c|}
\hline
Particle & $SU(2)_L$ & $T_3$         & $Y$  & $Q$  & $\gamma_5$-EV \\ \hline \hline
$e_L$, $\overline{(e_L)^\mathcal{C}}$\rule[0mm]{0mm}{4.5mm}
        & {\bf 2}   & $-\frac{1}{2}$ & $-1$ & $-1$ & $-1$ \\ \hline
$e_R$, $\overline{(e_R)^\mathcal{C}}$\rule[0mm]{0mm}{4.5mm}
        & {\bf 1}   & \hfill $0$     & $-2$ & $-1$ & $\ 1$  \\ \hline
$(e_L)^\mathcal{C}$, $\overline{e_L}$\rule[0mm]{0mm}{4.5mm}
        & {\bf 2}   & \hfill $\frac{1}{2}$  &\hfill $1$  & \hfill$1$  & $\ 1$ \\ \hline
$(e_R)^\mathcal{C}$, $\overline{e_R}$\rule[0mm]{0mm}{4.5mm}
        & {\bf 1}   & \hfill $0$     &\hfill $2$  &\hfill $1$  & $-1$ \\ \hline
\end{tabular}
\end{center}

\end{itemize}

% =============================================================================
\bibliographystyle{./apsrev}
\bibliography{./FlavourChanges}
% =============================================================================

\end{document}